\documentclass[superscriptaddress,prd,twocolumn,reprint,nofootinbib,preprintnumbers]{revtex4-2}

\usepackage[utf8]{inputenc}
\usepackage[final]{microtype}
\usepackage{lmodern}
\usepackage{exscale}
\usepackage[T1]{fontenc}

\usepackage{amsmath}
\usepackage{amssymb}
\usepackage{mathtools}

\usepackage[linktocpage]{hyperref}
\usepackage[dvipsnames,table]{xcolor}
\definecolor{Terracotta}{HTML}{e35336}
\hypersetup{
    colorlinks,
    linkcolor={Terracotta},
    citecolor={Terracotta},
    urlcolor={Terracotta}
}

\newcommand{\scr}{\scriptscriptstyle}

\usepackage{graphicx}
\graphicspath{ {./figures/} }

\usepackage{mathrsfs}

\newcommand{\MP}{M_{\scr \rm P}}
\newcommand{\RS}{r_{\rm \scr S}}

\makeatletter
\newcommand{\dalembertian}{\mathop{\mathpalette\dalembertian@\relax}}
\newcommand{\dalembertian@}[2]{%
  \begingroup
  \sbox\z@{$\m@th#1\square$}%
  \dimen0=\fontdimen8
    \ifx#1\displaystyle\textfont\else
    \ifx#1\textstyle\textfont\else
    \ifx#1\scriptstyle\scriptfont\else
    \scriptscriptfont\fi\fi\fi3
  \makebox[\wd\z@]{%
    \hbox to \ht\z@{%
      \vrule width \dimen0
      \kern-\dimen0
      \vbox to \ht\z@{
        \hrule height \dimen0 width \ht\z@
        \vss
        \hrule height 2\dimen0
      }%
      \kern-2.5\dimen0
      \vrule width 2.5\dimen0
    }%
  }%
  \endgroup
}
\makeatother

\usepackage{ulem}
\usepackage{cancel}

\makeatletter
\renewcommand{\fnum@figure}{{\bf Fig.~\thefigure}}
\makeatother

\usepackage{orcidlink}

\begin{document}

\preprint{IPARCOS-UCM-26-024}

\title{Cosmological properties of black-hole hair in the linearly
\vskip+1mm
coupled scalar-Gauss-Bonnet theory}

\author{Dra\v{z}en Glavan\,\orcidlink{0000-0002-1983-0448}\,}
\email[\tt email:\,]{glavan@fzu.cz}
\affiliation{CEICO, Institute of Physics of the Czech Academy of Sciences, 
Na Slovance 1999/2, 182 21 Prague 8, Czech Republic}

\author{Dar\'{i}o Jaramillo-Garrido\,\orcidlink{0009-0002-7123-4405}\,}
\email[\tt email:\,]{djaramil@ucm.es}
\affiliation{Departamento de F\'{i}sica Te\'{o}rica 
and Instituto de F\'{i}sica de Part\'{i}culas y del Cosmos
(IPARCOS-UCM), Universidad Complutense de Madrid, E-28040 Madrid, Spain}

\begin{abstract}
\noindent
We investigate the superhorizon behavior of scalar hair sourced by black holes in de Sitter spacetime in the linearly coupled shift-symmetric scalar-Gauss-Bonnet theory. Working in the test-field regime, we show that this hair exhibits both temporal and spatial growth on superhorizon scales. This growth is not a special consequence of the black hole, but instead follows from the dynamics of a minimally coupled massless scalar field in expanding de Sitter spacetime. Moreover, it is not even specific to black holes, but also arises for a point scalar charge in de Sitter, indicating that a scalarized black hole acts effectively as a localized subhorizon source of scalar perturbations. Backreaction, when important, first arises on subhorizon scales and does not by itself eliminate the superhorizon profile. The time-dependent scalar hair also carries a steady outward energy flux, which frames the test-field regime as a transient, and helps explain the difficulties encountered in attempts to construct self-consistent static solutions.
\end{abstract}

\maketitle

\section{Introduction}
\label{sec:Introduction}

Black holes in general relativity obey strong no-hair 
results~\cite{Bekenstein:1995un}, which in particular exclude 
nontrivial profiles of a canonical scalar field for 
asymptotically flat solutions under standard assumptions. 
Analogous no-hair results were established for static, 
spherically symmetric black holes in shift-symmetric 
Horndeski theories~\cite{Hui:2012qt}. However, these results 
do not apply to all shift-symmetric Horndeski theories, and an 
exception was identified in~\cite{Sotiriou:2013qea} for a 
particular subclass.

This exception arises when the scalar field couples linearly
to the Gauss-Bonnet invariant. In that case, the scalar-field
equation acquires a source term even in vacuum, so the
assumptions underlying the no-hair results no longer apply
(see also~\cite{Antoniou:2017acq} for a detailed study of coupling 
functions which evade the no-hair theorems).
Explicit asymptotically flat hairy black-hole solutions in
this linearly coupled shift-symmetric theory were constructed
perturbatively in the test-field regime
in~\cite{Sotiriou:2013qea}, and numerically as
self-consistent solutions in~\cite{Sotiriou:2014pfa}.
The resulting hair is secondary~\cite{Coleman:1991jf,Coleman:1991ku},
in the sense that it is not associated with an independent
asymptotic charge, but is instead fixed by the black-hole
parameters and the Gauss-Bonnet coupling.

For sufficiently small black-hole masses, the test-field
approximation breaks down because the scalar backreaction
can no longer be neglected, and the full coupled field
equations must be solved self-consistently. The numerical
analysis of~\cite{Sotiriou:2014pfa} showed that regular
black-hole solutions in this theory exist only above a
minimum mass. For the self-consistent solutions that were
constructed, deviations from the Schwarzschild geometry
remain comparatively small away from the immediate vicinity
of the black hole.

Subsequent work showed that this hair can form dynamically
from regular initial data, and also investigated the
stability properties of the corresponding black-hole
solutions~\cite{Ogawa:2015pea,Benkel:2016rlz}. This strengthens
the interpretation of the hairy solutions as physically
relevant configurations, rather than merely formal stationary
solutions. Moreover, the generalization to the more physically realistic Kerr black holes was also considered~\cite{Delgado:2020rev}.

The situation becomes more subtle once one asks whether these
hairy black-hole solutions can be embedded consistently in a
cosmological spacetime. Earlier numerical attempts to
construct static scalarized black holes with de Sitter
asymptotics found no solutions~\cite{Brihaye:2017wln,Bakopoulos:2018nui};
see also~\cite{Brihaye:2019gla}. Recent
work~\cite{Babichev:2024txe} proposed that this failure can be
traced to the cosmological horizon, which obstructs static
scalar hair for a spherically symmetric black hole in an
expanding de Sitter background.

In Ref.~\cite{Babichev:2024txe}, the scalar field was
allowed to depend linearly on time as well as radius, while still
yielding a stationary energy-momentum tensor because of shift
symmetry. An explicit solution for the scalar hair 
was constructed in static coordinates in the test field approximation.
It shows scalar hair that decays very slowly with distance from the black hole, and extending all the way to the cosmological horizons 
where it does not match the expected cosmologically rolling 
homogeneous and isotropic solution. The question of whether such a 
solution is physically reasonable was raised since it would suggest that a local black-hole configuration can qualitatively alter the large-distance cosmological behavior of the spacetime. It then becomes important to understand the physical origin of this unexpected feature.

Here we revisit the properties of black-hole scalar hair on 
cosmological scales. We show that its large-distance profile is 
controlled not by the local black-hole structure or by the 
scalarization mechanism specific to the linearly coupled
scalar-Gauss-Bonnet theory, but by the cosmological dynamics
of a minimally coupled massless scalar field. The
same dynamics underlies the generation of nearly
scale-invariant superhorizon scalar perturbations during
inflation~\cite{Parker:1968mv}.

Moreover, this profile is not unique to black holes. It was
obtained previously for a point scalar charge in de Sitter
space in~\cite{Burko:2002ge}, and later rederived
in~\cite{Akhmedov:2010ah,Glavan:2019yfc}. This indicates that,
on cosmological scales, a black hole is effectively
indistinguishable, as a source of the scalar field, from any
other static, spherically symmetric localized scalar charge. 
At those scales, the
information about the source enters only through the overall
amplitude of the profile, which is fixed locally by the
black-hole mass and the Gauss-Bonnet coupling.

From a cosmological perspective, a scalarized black hole acts as
a localized source of scalar perturbations well inside the
cosmological horizon. As the universe expands, these
perturbations are stretched to superhorizon scales and
amplified, giving rise to a nearly scale-invariant spectrum.
The resulting picture is simple: the black hole seeds the hair
locally, while cosmological expansion determines its asymptotic
form. We derive this picture explicitly for a point source in
de Sitter, and then show that the scalar profile of a black
hole in Schwarzschild-de Sitter spacetime follows the same
asymptotic pattern, with both systems also exhibiting an
outward energy flux associated with the time-dependent scalar
configuration.

To disentangle the different ingredients of the problem, we
consider three complementary settings: the Schwarzschild
spacetime in Sec.~\ref{sec: Schwarzschild spacetime}, a scalar
point source in de Sitter spacetime in
Sec.~\ref{sec: Scalar point source in de Sitter spacetime}, and
the Schwarzschild-de Sitter spacetime in
Sec.~\ref{Schwarzschild-de Sitter spacetime}. Each example is
analyzed in the test-field approximation, with particular
attention to the regime in which scalar backreaction becomes
important. We find that the first breakdown of the test-field
description occurs on subhorizon scales. Any reliable
conclusion about the cosmological-scale scalar-hair profile
therefore requires understanding the nonlinear dynamics near
the black hole.

\section{Preliminaries}
\label{sec: Preliminaries}

In this section we collect the basic equations of the theory and
introduce the invariant diagnostic that will be used throughout
the paper to estimate when the test-field approximation is
expected to break down.

We consider the linearly coupled shift-symmetric 
scalar-Gauss-Bonnet (sGB) theory, with action
\begin{equation}
S[g_{\mu\nu},\Phi] = \!\! \int\! {\rm d}^4x \, \sqrt{-g} \, \biggl[ \frac{\MP^2}{2} (R-2\Lambda)
	+ X
	+ \alpha \Phi \mathcal{G}
	\biggr] 
	\, ,
\label{OriginalAction}
\end{equation}
where~$X\!=\! - \frac{1}{2} g^{\mu\nu} \partial_\mu \Phi \partial_\nu \Phi$ is the scalar kinetic term,~$\Lambda$
is the cosmological constant,~$\MP \!=\! 1/\sqrt{8\pi G_{\scr N}}$ is the reduced Planck mass,
and~$\mathcal{G} \!=\! R^2 \!-\! 4 R^{\mu\nu}R_{\mu\nu} \!+\! R^{\mu\nu\rho\sigma}R_{\mu\nu\rho\sigma}$
is the Gauss-Bonnet invariant.

The scalar equation of motion is
\begin{equation}
\dalembertian \Phi = \frac{1}{\sqrt{-g} } 
	\partial_\mu \big( \sqrt{-g} \, g^{\mu\nu} \partial_\nu \Phi \big)
	= - \alpha \mathcal{G} \, ,
\label{eom}
\end{equation}
while the energy-momentum tensor is
\begin{equation}
\mathcal{T}_{\mu\nu}
    =
    \nabla_\mu\Phi \nabla_\nu\Phi
    +
    X g_{\mu\nu}
    -
    8 \alpha P_{\mu\rho\nu\sigma} \nabla^\rho \nabla^\sigma \Phi \, ,
\end{equation}
where~$P_{\mu\nu\rho\sigma} \!=\! R_{\mu\nu\rho\sigma} 
\!-\! 4 g_{\mu][\rho} R_{\sigma][\nu}
\!+\! R g_{\mu[\rho} g_{\sigma]\nu} $
is proportional to the double-dual of the Riemann tensor,
$P_{\mu\nu\rho\sigma} \!=\! (-1) \!\times\! \frac{1}{4} \epsilon_{\mu\nu\alpha\beta} \epsilon_{\rho\sigma\gamma\delta} R^{\alpha\beta\gamma\delta}$.
We will frequently use the trace of the energy-momentum tensor,
\begin{equation}\label{eq: EMT trace}
\mathcal{T} = 2X + 8\alpha G_{\mu\nu} \nabla^\mu \nabla^\nu \Phi \, ,
\end{equation}
where~$G_{\mu\nu} \!=\! R_{\mu\nu} \!-\! \frac{1}{2} g_{\mu\nu} R$
is the Einstein tensor.

Since we will consider the Schwarzschild, de Sitter, and
Schwarzschild-de Sitter spacetimes, it is useful to introduce
a simple invariant estimate of the size of scalar backreaction.
We do so by comparing the trace of the scalar energy-momentum
tensor to the local curvature scale set by the Kretschmann
invariant,
\begin{equation}
\mathcal{K} = R^{\mu\nu\rho\sigma} R_{\mu\nu\rho\sigma} \, .
\label{bDiagnostic}
\end{equation}
Accordingly, we define
\begin{equation}
b \equiv \Big|
\frac{\mathcal{T}}{\MP^2 \sqrt{\mathcal{K}}}
\Big|\, .
\label{bDef}
\end{equation}
This quantity does not replace a full solution of the coupled
field equations, but it provides a convenient invariant
diagnostic for identifying the regime in which the test-field
approximation is expected to fail.

\section{Schwarzschild spacetime}
\label{sec: Schwarzschild spacetime}

We begin with the Schwarzschild spacetime, summarizing the
derivation of~\cite{Sotiriou:2013qea} in order to collect the
results that will be useful later and to establish the picture
that we expect to remain valid on subhorizon scales in the
cosmological setting, in the regime where the two horizons 
are widely separated that will be considered in the
following sections.

The line element in Schwarzschild coordinates is
\begin{equation}
{\rm d}s^2 = - f(r) {\rm d}\tau^2 + \frac{{\rm d}r^2}{f(r)} + r^2{\rm d}\Omega^2 \, .
\label{SchLineElement}
\end{equation}
where~$f(r)\!=\! 1 \!-\! \RS/r$,
and where~$r_{\rm \scr S}\!=\!2GM$ is the Schwarzschild radius
related to the black hole mass~$M$. In this spacetime, one 
has~$\mathcal{G} \!=\! \mathcal{K}
    \!=\! 12 \RS^2/r^6$. Assuming a radially dependent scalar
profile, $\Phi\!=\!\Phi(r)$, the equation of
motion~(\ref{eom}) reduces to
\begin{equation}
\frac{{\rm d}}{{\rm d}r} \Big( r^2 f(r) \frac{{\rm d}\Phi}{{\rm d}r} \Big) 
    = - \frac{12\alpha \RS^2}{r^4}
\, .
\end{equation}
Integrating once gives
\begin{equation}
\frac{{\rm d}\Phi}{{\rm d}r} =
    \frac{4\alpha r_{\scr \rm S}^2}{r^5 f(r)}
    -
    \frac{\mathcal{C}}{r^2f(r)}
    \, ,
\end{equation}
where~$\mathcal{C}$ is an integration constant. Requiring the
scalar gradient to remain finite at the black-hole horizon fixes
it to
\begin{equation}
\mathcal{C} = \frac{4\alpha}{\RS} \, .
\end{equation}
This shows that the scalar hair is secondary: the integration constant 
is not an independent scalar charge, but is instead fixed by the coupling
constant and the black-hole mass.

Consequently, the scalar hair profile, up to an irrelevant integration
constant, is
\begin{equation}
\Phi(r) = \frac{4\alpha}{\RS r}
    \bigg[ 1 + \frac{\RS}{2r} + \frac{\RS^2}{3r^2} \bigg]
    \, .
\label{SchProfile}
\end{equation}
An important point is that this profile remains finite at the
horizon even in Schwarzschild coordinates. Thus, in this
particular case there is no need to invoke horizon-regular
coordinates in order to establish regularity of the scalar
field. Nevertheless, it will be useful to introduce such
coordinates already here, since they provide a natural basis
for comparison with the more complicated cases to follow. A
convenient choice is given by Painlev\'{e}-Gullstrand
coordinates~\cite{Gaur:2022hap}, obtained from Schwarzschild
coordinates through the time transformation,
\begin{equation}
\tau = t + \int\! \frac{{\rm d}r}{f(r)} \sqrt{ 1\!-\! f(r) } \, .
\end{equation}
This puts the line element~(\ref{SchLineElement}) in the 
manifestly nonsingular form,
\begin{equation}
{\rm d}s^2 = - {\rm d}t^2 + \Big( {\rm d}r - \sqrt{1 \!-\! f(r)} \, {\rm d}t \Big)^{\!2} + r^2 {\rm d}\Omega^2 \, .
\label{SchPainleveGullstrand}
\end{equation}
Because the scalar hair profile~(\ref{SchProfile}) is static, 
it is unchanged by this coordinate transformation and remains 
a function of~$r$ only.

\begin{figure}[h!]
\center
\includegraphics[width=8.5cm]{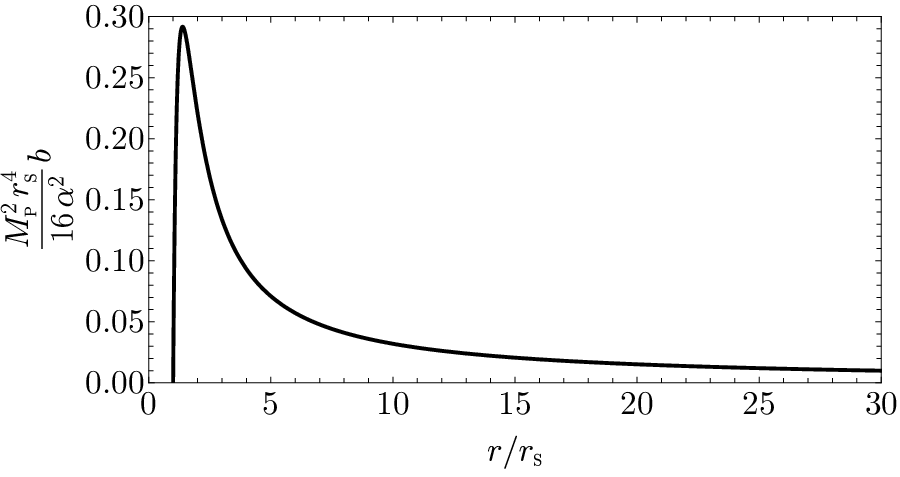}
\vspace{-3mm}
\caption{Strength of the backreaction of the scalar field on the
Schwarzschild background. The backreaction first becomes important 
close to the black hole.}
\label{SchBackreactionPlot}
\end{figure}

Having obtained the scalar hair profile, we can compute the
trace of the energy-momentum tensor. In the Schwarzschild
background it receives no direct contribution from the
Gauss-Bonnet coupling, but only from the scalar kinetic term,
\begin{equation}
\mathcal{T} = - \frac{16\alpha^2}{r_{\scr \rm S}^2}
	\frac{ f(1\!-\!f)^4 }{r_{\scr \rm S}^4} 
        \Big[ f^2 + 3(1\!-\!f) \Big]^{2}
	\, .
\end{equation}
This should be compared with the Kretschmann invariant,
\begin{equation}
\mathcal{K} = \frac{12 r_{\scr \rm S}^2 }{r^6} \, ,
\end{equation}
which sets the local curvature scale of the Schwarzschild 
geometry. The corresponding backreaction measure~(\ref{bDiagnostic}) 
is therefore
\begin{equation}
b
	=
	\frac{16\alpha^2}{\MP^2 r_{\scr \rm S}^4} 
    \frac{f(1\!-\!f) }{\sqrt{12} }
	\Big[ f^2 \!+\! 3(1\!-\!f) \Big]^{2}
	\, .
\label{bSch}
\end{equation}
Its behavior is shown in Fig.~\ref{SchBackreactionPlot}. The
backreaction becomes appreciable first close to the black-hole horizon, while 
remaining small both at the horizon and at large distances.

An important consequence of Eq.~(\ref{bSch}) is that the
backreaction becomes stronger for smaller black holes. Since the
backreaction measure scales inversely with the fourth power of
the Schwarzschild radius, decreasing~$r_{\scr \rm S}$ drives the
system more quickly toward the breakdown of the test-field
approximation. This is precisely why solving the full coupled
field equations becomes necessary for sufficiently small black
holes. Such self-consistent solutions were constructed
numerically in~\cite{Sotiriou:2014pfa}, where this behavior was
shown to imply a minimum allowed black-hole mass in the theory.

\section{Point source in de Sitter spacetime}
\label{sec: Scalar point source in de Sitter spacetime}

A point scalar charge in de Sitter spacetime provides the
simplest setting in which to isolate the cosmological behavior
of a localized scalar source, without the additional near-source
structure associated with a black hole. It therefore allows us
to distinguish the generic cosmological properties of the
large-distance scalar profile from those tied specifically to
black-hole scalarization.

We adopt conformal cosmological coordinates for expanding
de Sitter spacetime,
\begin{equation}
{\rm d}s^2 = a^2(\eta) \big( - {\rm d}\eta^2 + {\rm d} \vec{x}^{\,2} \big) \, ,
\label{ConformalCoordinates}
\end{equation}
$\eta \!\in\!(-\infty,0)$ is conformal time,
$a(\eta) \!=\! -1/(H\eta)$ is the scale factor, and
$H\!=\!\sqrt{\Lambda/3}$ is the Hubble rate.

We consider the scalar theory~(\ref{OriginalAction}) on this
background. The scalar is sourced both by the Gauss-Bonnet
invariant of de Sitter spacetime,~$\mathcal{G} \!=\! 24H^4$,
and by a scalar point charge placed at the origin of the
comoving coordinate system,\footnote{The scalar
charge~$\lambda$ used in~\cite{Glavan:2019yfc} is related to
the one we use here as~$\lambda\!=\!-4\pi \beta$.}
\begin{equation}
{\dalembertian} \Phi
=
\!- \frac{1}{a^2}\Big[ \partial_\eta^2
    + 2aH \partial_\eta
    - \nabla^2 \Big]\Phi
=\!
-24\alpha H^4 - \frac{4\pi \beta \delta^3(\vec{x})}{a^3}
\, .
\end{equation}

Because the equation of motion is linear, the full solution can
be decomposed into the homogeneous part sourced by the de
Sitter background and the part sourced by the point charge. The
former is removed by the time-dependent shift,
\begin{equation}
\Phi(\eta,x)
    = 
    8\alpha H^2 \ln(a)
    + \phi(\eta,x)
    \, ,
\label{PointSourcePhiCosmo}
\end{equation}
where~$x \!=\! \| \vec{x} \|$,
so that the remaining field satisfies
\begin{equation}
\Big[ \partial_\eta^2 + 2aH\partial_\eta - \nabla^2 \Big]\phi
=
\frac{4\pi\beta \delta^3(\vec{x})}{a} \, .
\label{phiEq}
\end{equation}
This is precisely the equation studied
in~\cite{Burko:2002ge,Akhmedov:2010ah,Glavan:2019yfc} for a
minimally coupled massless scalar field sourced by a point
charge in de Sitter spacetime.\footnote{For clarity,
Refs.~\cite{Glavan:2019yfc,Glavan:2021adm} concern a minimally
coupled massless scalar field sourced by a point charge in de
Sitter spacetime (and quantum-gravitational corrections to its
evolution), rather than the Brans-Dicke theory as stated
in~\cite{Babichev:2024txe,Babichev:2025ric}.}

The solution to Eq.~(\ref{phiEq}) was obtained
in~\cite{Burko:2002ge,Akhmedov:2010ah,Glavan:2019yfc} and reads
\begin{equation}
\phi(\eta,x) =
    \frac{\beta}{ a x}
    + \beta H \ln\!\Big( \frac{a}{1 + aHx} \Big)
\, .
\label{PointSourcePhiRadial}
\end{equation}
An important feature of this solution is that it breaks the
dilation symmetry of de Sitter spacetime and grows
logarithmically on superhorizon scales,
\begin{equation}
\phi(\eta,x) \xrightarrow{ax \gg 1/H} 
    - \beta H \bigg[ \ln(Hx) - \frac{1}{2(Hax)^2}
    + \mathcal{O}(Hax)^{-3} \bigg]
    \, .
\label{phiSuper}
\end{equation}
At first sight, the absence of decay at large distances might be
taken to signal a tension with the cosmological horizon. As we will show, however,
this behavior is naturally understood in terms of the
cosmological dynamics of a massless, minimally coupled scalar
field. This dynamics is familiar from inflationary cosmology,
where subhorizon scalar perturbations are stretched to
superhorizon scales by the expansion and are amplified in the
process, yielding a nearly scale-invariant spectrum of
primordial scalar perturbations~\cite{Planck:2018jri}. 
The same mechanism underlies the superhorizon scalar profile
sourced by the point charge.

\subsection{Spectrum}
Because the scalar equation of motion is linear, the
homogeneous contribution induced by the Gauss-Bonnet coupling
can be separated from the part sourced by the scalar point
charge. It is therefore sufficient to focus here on the latter,
since the homogeneous contribution affects only the zero mode.

To see explicitly how this superhorizon behavior arises, we
consider the point charge to be switched on at a finite
conformal time~$\eta_0$, at which we take~$a(\eta_0)\!=\!1$.
The equation of motion then becomes
\begin{equation}
\Big[ \partial_\eta^2  +2 aH \partial_\eta - \nabla^2 \Big] \phi
	= \theta(\eta \!-\! \eta_0) 
    \frac{4\pi \beta \delta^3(\vec{x})}{a} \, ,
\label{InstantaneousTurningOnEq}
\end{equation}
where~$\theta(z)$ is the Heaviside step function.
This equation was solved in position space
in~\cite{Glavan:2019yfc} using the Green's function method,
yielding
\begin{equation}
\phi(\eta,x) = \theta \big( \eta \!-\! \eta_0 \!-\! x \big)
	\bigg[ \frac{\beta}{ a x}
		+ \beta H \ln \! \Big( \frac{a}{1 + aHx} \Big)
        \bigg]
	\, .
\end{equation}
This makes explicit that the logarithmically growing
large-distance profile is established causally, rather than
being present instantaneously throughout de Sitter spacetime,
including beyond the cosmological horizon.

A more detailed physical picture of this mechanism is obtained
by examining the scalar profile in momentum space. Its Fourier
transform,
\begin{equation}
\varphi(\eta,k) = \int \! {\rm d}^3x \, 
    e^{-i\vec{k} \cdot \vec{x}}
\, \phi(\eta,\vec{x}) \, ,
\end{equation}
depends only on~$k\!=\!\|\vec{k}\|$ owing to spherical symmetry,
and satisfies
\begin{equation}
\varphi'' + 2 a H \varphi' + k^2 \varphi
	= \theta(\eta \!-\! \eta_0) \frac{4\pi\beta}{a} \, ,
\end{equation}
with primes denoting derivatives with respect to conformal
time.

\begin{figure}[h!]
\center
\includegraphics[width=8.5cm]{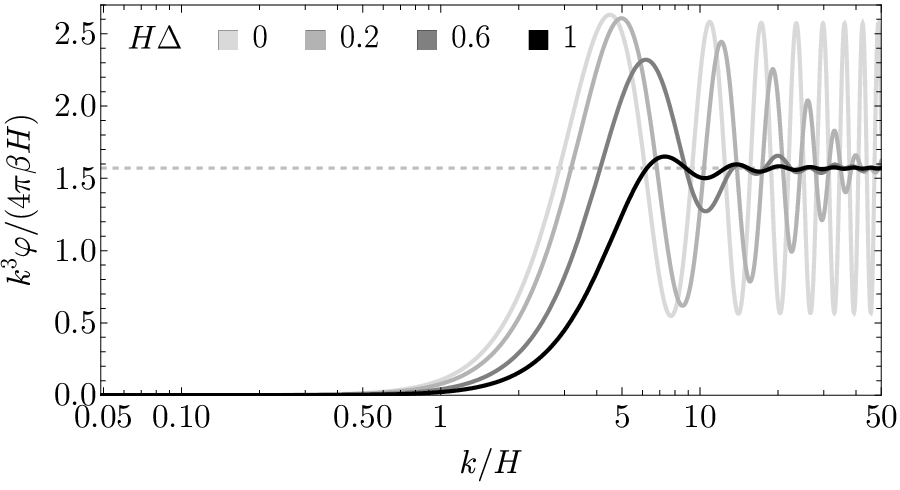}
\vspace{-3mm}
\caption{Late-time spectra of the scalar field sourced by a point
charge in de Sitter spacetime, for source profiles whose
switching on begins at conformal time~$\eta_0$, 
at which~$a(\eta_0)\!=\!1$. The curve with~$\Delta\!=\!0$
corresponds to the instantaneous switching on described by
Eq.~(\ref{InstantaneousTurningOnEq}). The other three curves
correspond to a finite switching-on interval, implemented by
replacing the step function
in~(\ref{InstantaneousTurningOnEq}) according to
$\theta(\eta \!-\! \eta_0) \!\to\!
\Theta\big[(\eta\!-\!\eta_0)/\Delta\big]$, where
$\Theta(z)$ is taken to be the {\it smoothstep} function,
$\Theta(z) \!\equiv\! \theta(z)\big[1 \!-\!
\theta(1\!-\!z)(1\!+\!2z)(1\!-\!z)^2\big]$. The
parameter~$\Delta$ sets the switching-on duration.
Only modes that were subhorizon when the
source was switched on, $k/H \!>\! 1$, become populated on
superhorizon scales. The oscillations superimposed on the
scale-invariant part of the spectrum are a consequence of
abrupt switching and are smoothed out when the source is turned
on more gradually.}
\label{SpectrumFigure}
\end{figure}

The solution satisfying the initial 
conditions~$\varphi(\eta_0,k)\!=\!0$
and~$\varphi'(\eta_0,k) \!=\! 0$, which correspond to the
instantaneous switching on of the source, is obtained
analytically as
\begin{align}
\varphi(\eta,k) ={}&
    \frac{4\pi\beta H}{k^3}
    \bigg\{
    \frac{k}{Ha}
    -
    \Big[ \cos\!\Big( \frac{k}{aH} \Big)
        + \frac{k}{aH}\sin\!\Big( \frac{k}{aH} \Big)
    \Big]
\nonumber \\
&
	\times
    \Big[
    \sin\!\Big( \frac{k}{H} \Big)
    - {\rm Si} \Big( \frac{k}{H} \Big)
    + {\rm Si}\Big( \frac{k}{Ha} \Big)
    \Big]
\nonumber \\
& \hspace{-0.7cm}
    +
    \Big[ \sin\!\Big( \frac{k}{aH} \Big)
        - \frac{k}{aH}\cos\!\Big( \frac{k}{aH} \Big)
    \Big]
\nonumber \\
&
	\times
    \Big[
    \cos\!\Big( \frac{k}{H} \Big)
    - {\rm Ci} \Big( \frac{k}{H} \Big)
    + {\rm Ci}\Big( \frac{k}{Ha} \Big)
    \Big]
    \bigg\}
    \, ,
\end{align}
where the sine- and cosine-integral functions are defined
by~${\rm Si}(x) \!=\! \int_0^x \! {\rm d}y\, \sin(y)/y $
and~${\rm Ci}(x) \!=\! - \int_x^\infty \! {\rm d}y\, \cos(y)/y$.
The late-time limit of the solution takes the particularly
simple form,
\begin{equation}
\varphi(\eta,k) \xrightarrow{a \to \infty}
	\frac{4\pi\beta H}{k^3}
    \bigg[ {\rm Si} \Big( \frac{k}{H} \Big)
    - \sin\!\Big( \frac{k}{H} \Big) \bigg] \, ,
\end{equation}
and the corresponding spectrum is shown in
Fig.~\ref{SpectrumFigure}, together with spectra obtained for
slower switching-on of the source.

This spectrum makes the physical picture particularly clear.
Only those modes that were subhorizon when the source was
switched on, namely~$k\!>\!H$, become populated. By contrast,
modes that were already superhorizon at that time,
namely~$k\!<\!H$, remain unpopulated. The point source therefore
first seeds scalar perturbations on subhorizon scales, after
which cosmological evolution takes over: the modes are
stretched to superhorizon scales and amplified in the familiar
way for a massless, minimally coupled scalar field. The result
is a scale-invariant spectrum on superhorizon scales.
This supports the interpretation that the superhorizon scalar
profile is a consequence of cosmological scalar-field dynamics
rather than of any detailed properties of the stationary,
spherically symmetric source.

\subsection{Painlev\'{e}-Gullstrand coordinates}
For later comparison with the Schwarzschild-de Sitter case, it
is useful to rewrite the point-source solution in
Painlev\'{e}-Gullstrand coordinates~\cite{Gaur:2022hap}. These
are obtained by adopting the physical distance as the radial
coordinate,~$r \!=\! ax$, and the physical time defined
by~${\rm d}t\!=\!a{\rm d}\eta$. The line element then reads
\begin{equation}
{\rm d}s^2 = - {\rm d}t^2 + \big( {\rm d}r - H r {\rm d}t \big)^{\!2} + r^2 {\rm d}\Omega^2 \, .
\label{dS-PainleveGullstrand}
\end{equation}

Since the scalar field is a spacetime scalar, its profile in
these coordinates is obtained by rewriting the previous
solution as
\begin{equation}
\Phi(t,r) = ( 8 \alpha H \!+\! \beta ) H^2 t
    - \beta H \ln(1 \!+\! Hr)
    + \frac{\beta}{r} 
    \, .
\label{eq: field profile in PG coordinates}
\end{equation}
At superhorizon distances,
\begin{equation}
\Phi(t,r) \xrightarrow{r \gg 1/H} 
    ( 8 \alpha H \!+\! \beta ) H^2 t
    - \beta H \ln(Hr)
    \, ,
\label{dSasymptotic}
\end{equation}
it grows logarithmically with physical distance and linearly in
time. This is the same large-distance behavior that will later
be recovered for the Schwarzschild-de Sitter spacetime.

The scalar kinetic term, however, depends only on the radial
coordinate,
\begin{equation}
X = \frac{H^4}{2}
    \bigg[
    64 \alpha^2 H^2
    -
    \frac{16\alpha\beta H}{Hr(1 \!+\! Hr)}
    -
    \frac{\beta^2 ( 1 \!+\! H^2r^2 ) }{H^4 r^4}
    \bigg]
    \, .
\end{equation}
Although the scalar profile itself grows on superhorizon
scales, the scalar kinetic term controlling the
energy-momentum tensor remains much better behaved,
\begin{equation}
X \xrightarrow{r \gg 1/H} \frac{H^4}{2}
    \bigg[
    64 \alpha^2 H^2
    -
    \frac{\beta ( 16\alpha H + \beta ) }{H^2r^2}
    \bigg]
    \, .
\end{equation}
In particular, it asymptotes to a constant, with deviations
falling off as the inverse square of the physical distance.

\subsection{Backreaction}
The scalar profile~(\ref{dSasymptotic}) grows logarithmically with
physical distance on superhorizon scales, in addition to
growing linearly in cosmological time. This, however, does not
by itself imply large backreaction or an inconsistency of the
de Sitter background.

Backreaction is controlled not by the scalar profile itself,
but by the energy-momentum tensor, whose source-dependent part
decreases with distance. The logarithmic
growth of the scalar profile is therefore not, by itself,
evidence for a breakdown of the test-field approximation. This
can be made explicit by comparing the Kretschmann invariant,
which is constant in de Sitter,~$\mathcal{K} \!=\!
24 H^4$, to the trace of the energy-momentum tensor,
\begin{equation}
\mathcal{T} =
    H^4
    \bigg[
    640 \alpha^2 H^2
    -
    \frac{16\alpha\beta H}{Hr(1 \!+\! Hr)}
    -
    \frac{\beta^2 (1 \!+\! H^2r^2 ) }{H^4r^4}
    \bigg]
    \, .
\label{dST}
\end{equation}
The three terms in~(\ref{dST}) have distinct origins: the
first arises from the coupling to the Gauss-Bonnet term of
the de Sitter background, the third from the scalar point
source, and the middle term is the cross-term between the two.

\begin{figure}[h!]
\center
\includegraphics[width=8.5cm]{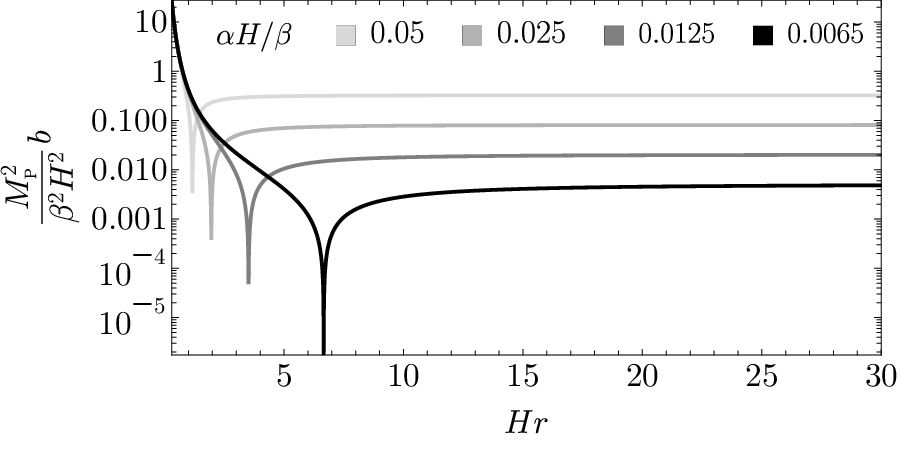}
\vspace{-3mm}
\caption{Strength of the backreaction induced by a point scalar
charge on the de Sitter background. The source-induced
backreaction is strongest on subhorizon scales and decreases
with distance, even though the scalar profile itself grows on
superhorizon scales.}
\label{dSBackreactionPlot}
\end{figure}

The corresponding invariant backreaction diagnostic is
\begin{equation}
b = \frac{\beta^2 H^2}{2\sqrt{6} \MP^2} \Bigg|
    640 \Big(\frac{\alpha H}{\beta} \Big)^{\!2}
    \! -
    \frac{16(\alpha H/\beta)}{Hr(1 \!+\! Hr)}
    -
    \frac{1 \!+\! H^2 r^2 }{H^4r^4}
    \Bigg|
    \, .
\end{equation}
Its behavior is shown in Fig.~\ref{dSBackreactionPlot}.
Near the source, that is on subhorizon scales,
\begin{equation}
b \xrightarrow{r \ll 1/H}
    \frac{\beta^2 H^2}{2\sqrt{6} \MP^2} \bigg[
    \frac{1}{H^4r^4}
    +
    \frac{1}{H^2r^2}
    \bigg]
    \, ,
\end{equation}
so the source-induced contribution grows rapidly as the source is
approached. By contrast, on superhorizon scales,
\begin{equation}
b \xrightarrow{r \gg 1/H}
    \frac{\beta^2 H^2}{2\sqrt{6} \MP^2} \bigg[
    640\Big(\frac{\alpha H}{\beta}\Big)^2
    -
    \frac{16(\alpha H /\beta)+1}{H^2r^2}
    \bigg]
    \, .
\end{equation}
Thus, even though the scalar profile itself grows
logarithmically at large distances, the source-dependent part
of the backreaction becomes weaker with increasing distance.

An important consequence is that if the source-induced
backreaction became appreciable on cosmological scales, then it
would necessarily have been much larger on subhorizon scales.
The first breakdown of the test-field approximation therefore
occurs near the source rather than as a cosmological-scale
obstruction, and must be understood there before drawing
conclusions about cosmological scales.

\subsection{Scalar radiation}
A point scalar charge in de Sitter spacetime also carries a
nonvanishing radial energy flux, naturally interpreted as
monopole scalar radiation~\cite{Burko:2002ge}. To exhibit this
explicitly, it is convenient to rewrite the point-source
solution in static coordinates. These are related to the
conformal cosmological coordinates by
\begin{equation}
\tau = - \frac{1}{H} \ln
    \bigg[ \frac{\sqrt{ 1 \!-\! H^2 a^2(\eta) x^2 }}{a(\eta)} \,
    \bigg] \, ,
\qquad
r = a(\eta) x \, ,
\end{equation}
so that the line element becomes
\begin{equation}
{\rm d}s^2 = - h(r) {\rm d}\tau^2 + \frac{{\rm d}r^2}{h(r)}
    + r^2 {\rm d}\Omega^2 \, ,
\end{equation}
where~$h(r) \!=\! 1 \!-\! H^2 r^2$. These coordinates cover 
only the static patch, that is, the region inside the 
cosmological horizon~$r \!<\! 1/H$, where the time 
direction corresponds to the Killing vector. This makes 
these coordinates well-suited for describing the associated 
stationary energy flux.

The total inward energy flux through a sphere of radius~$r$ 
is defined by~$\dot{M} \!=\! 4\pi r^2 T^r{}_\tau$.
To evaluate it, we
rewrite the scalar profile~(\ref{PointSourcePhiCosmo}) in
static coordinates,
\begin{align}
\Phi(\tau,r)
    ={}&
    (8\alpha H \!+\! \beta) H^2 \tau
    +
    4\alpha H^2 \ln\! \big( 1 \!-\! H^2r^2 \big)
\nonumber \\
&
    +
    \beta H \ln\!\bigg[
        \frac{\sqrt{1 \!-\! H^2r^2}}{1 \!+\! Hr} \, \bigg]
    +
    \frac{\beta}{r}
    \, .
\end{align}
The profile is singular at the cosmological horizon in these
coordinates, but this is only an artifact of the static patch.
A direct computation then gives the constant negative flux,
\begin{equation}
\dot{M} = - 4\pi \beta (\beta \!+\! 8\alpha H ) H^2 \, .
\label{ACCdS}
\end{equation}
This is a purely de Sitter effect,\footnote{While this effect of scalar monopole radiation 
is not observed in flat space
in 3+1 dimensions, it is found in lower 2+1 and 1+1 dimensions~\cite{Burko:2002gf}.}
which vanishes in the flat-space limit~$H\!\to\!0$. In static
coordinates, it describes a stationary outward energy flux and
admits the natural interpretation of monopole scalar radiation,
as discussed in Ref.~\cite{Burko:2002ge}; see
also~\cite{Quinn:2000wa,Poisson:2003nc}.

This effect suggests that the point source cannot support the
same stationary scalar profile indefinitely. The outward flux
indicates that the system must eventually evolve beyond the
test-field description, as the source radiates away its entire
mass~\cite{Burko:2002ge}. However, the detailed late-time fate
of the source and of the scalar profile it supports is not yet
fully understood~\cite{Akhmedov:2010ah}.

\section{Schwarzschild-de Sitter spacetime}
\label{Schwarzschild-de Sitter spacetime}

We now turn to the Schwarzschild-de Sitter
spacetime, which combines the local black-hole structure of
the Schwarzschild solution with the large-scale cosmological
structure of de Sitter space. This is the geometry in which
the two mechanisms isolated in the previous sections coexist
and can be studied together.

The Schwarzschild-de Sitter line element in Kottler 
coordinates is
\begin{equation}
{\rm d}s^2 = - F(r) {\rm d}\tau^2 + \frac{{\rm d}r^2}{F(r)} + r^2 {\rm d}\Omega^2 \, ,
\label{KottlerCoordinates}
\end{equation}
where~$F(r) \!=\! 1 \!-\! r_{\scr \rm S}/r \!-\! H^2 r^2$. 
The black-hole horizon and the cosmological horizon, 
respectively, are located at
\begin{align}
r_{\scr \rm BH} ={}&
    \frac{2}{\sqrt{3} H} \sin \! \bigg[ \frac{1}{3}
    \arcsin \! \Big( \frac{3\sqrt{3} }{2} Hr_{\scr \rm S} \Big) 
    \bigg]
    \, ,
\\
r_{\scr \rm C} ={}&
    \frac{1}{H} \cos \! \bigg[ \frac{1}{3}
    \arcsin \! \Big(\frac{ 3\sqrt{3} }{2} Hr_{\scr \rm S} \Big) 
    \bigg]
    -
    \frac{r_{\scr \rm BH}}{2}
    \, .
\end{align}

We focus on the regime in which the black-hole and cosmological
horizons are widely separated, so that~$H r_{\scr \rm S}
\!\ll\! 1$. In this limit, the horizon radii are well-approximated by the parameters appearing in the line element,
\begin{align}
r_{\rm \scr BH} ={}& 
    r_{\scr \rm S}
    \Big[ 1 + (H r_{\scr \rm S})^2 
        + \mathcal{O}(H r_{\scr \rm S})^4 \Big] \, ,
\label{BHhorizon}
\\
r_{\scr \rm C} ={}& 
    \frac{1}{H} 
    \Big[ 1 - \tfrac{1}{2}Hr_{\scr \rm S} 
        - \tfrac{3}{8} (H r_{\scr \rm S})^2
	+ \mathcal{O}(H r_{\scr \rm S})^3 \Big]
	\, .
\label{Chorizon}
\end{align}
%

\subsection{Integrating equation of motion}
Following the static-coordinate strategy of~\cite{Babichev:2024txe},
we adopt the ansatz
\begin{equation}
\Phi(\tau, r) = \mathcal{Q} H^2 \tau + \varphi(r) \, ,
\end{equation}
where the first term is a homogeneous solution, while the
radial function~$\varphi(r)$ captures the part sourced by the
Gauss-Bonnet invariant,~$\mathcal{G} \!=\! 24 H^4 \!+\! 12\RS^2/r^6 $.
The equation of motion then reduces to
\begin{equation}
\partial_r
	\Big( r^2 F(r) \partial_r \varphi \Big)
	=
	- \alpha \bigg( \frac{12 r_{\scr \rm S}^2}{r^4} \!+\! 24 H^4 r^2 \bigg)
	\, ,
\end{equation}

The equation can then be integrated once, yielding
\begin{equation}
\partial_r \varphi = 
    \frac{1}{r^2 F(r)}
    \Big(
    4\alpha F'(r) \big[ 1 \!-\! F(r) \big]
	-
	\mathcal{C}
    \Big)
	\, ,
\end{equation}
where~$\mathcal{C}$ is an integration constant. A second
integration would produce only an additive constant in the
scalar field, which is physically irrelevant because of shift
symmetry.

The two physically relevant constants are therefore
$\mathcal{Q}$ and~$\mathcal{C}$. They are constrained by
requiring finiteness of the scalar kinetic term,
\begin{equation}
X = \frac{\mathcal{Q}^2 H^4}{2F(r)}
    -
    \frac{1}{2r^4 F(r)}
    \Big(
    4\alpha F'(r) \big[ 1 \!-\! F(r) \big]
	-
	\mathcal{C}
    \Big)^{\!2}
	\, ,
\end{equation}
at the two horizons. Demanding regularity at the black-hole
horizon~(\ref{BHhorizon}) and at the cosmological
horizon~(\ref{Chorizon}) yields, respectively,
\begin{align}
   (Hr_{\scr \rm BH})^4 \mathcal{Q}^2
    ={}&
    \big[
        4\alpha F'(r_{\scr \rm BH})
	    - \mathcal{C}
       \big]^2
    \, ,
\label{condition1}
\\
    (Hr_{\scr \rm C})^4 \mathcal{Q}^2
    ={}&
    \big[ 4\alpha F'(r_{\scr \rm C})
	    - \mathcal{C}
        \big]^2
    \, .
\label{condition2}
\end{align}
At this stage, however, these conditions still admit four
algebraic branches,
\begin{align}
\mathcal{C} ={}&
    4\alpha
    \frac{ r_{\scr \rm C}^2 F'(r_{\scr \rm BH})
            \mp r_{\scr \rm BH}^2 F'(r_{\scr \rm C}) }
        { r_{\scr \rm C}^2 \mp r_{\scr \rm BH}^2 }
        \, ,
\\
|\mathcal{Q}| ={}&
    4\alpha \frac{ F'(r_{\scr \rm BH})
            - F'(r_{\scr \rm C}) }
        { H^2 ( r_{\scr \rm C}^2 \mp r_{\scr \rm BH}^2 ) }
    \, .
\end{align}

This branch ambiguity is not most naturally resolved in Kottler
coordinates, since they are singular at the cosmological
horizon, just as static coordinates are in pure de Sitter
space. The behavior of the scalar profile across the horizons,
and hence the selection of the appropriate branch, are more
naturally analyzed in coordinates that remain regular at both
horizons. We now turn to such a description.

\subsection{Painlev\'{e}-Gullstrand coordinates}
A horizon-regular representation of the Schwarzschild-de
Sitter spacetime is given by Painlev\'{e}-Gullstrand
coordinates~\cite{Gaur:2022hap}. The transformation
$\tau \!=\! t \!+\! w(r)$, with
\begin{equation}
w(r)= \int\! \frac{{\rm d}r}{F(r)} \sqrt{1 \!-\! F(r)}
\end{equation}
brings the line element to the form
\begin{equation}
{\rm d}s^2 = - {\rm d}t^2
    + \Big({\rm d}r - \sqrt{1\!-\!F(r)} \,{\rm d}t \Big)^{\!2} 
    + r^2{\rm d}\Omega^2 .
\end{equation}
It reduces to the Schwarzschild line element in
Painlev\'{e}-Gullstrand form~(\ref{SchPainleveGullstrand})
in the limit~$H\!\to\!0$, and to the de Sitter line element in
Painlev\'{e}-Gullstrand form~(\ref{dS-PainleveGullstrand})
in the limit~$r_{\scr \rm S}\!\to\!0$.

The full scalar profile in these coordinates reads
\begin{equation}
\Phi(t, r) = \mathcal{Q} H^2 t
    + \mathcal{Q} H^2 w(r)  + \varphi(r) \, .
\end{equation}
It is more convenient, however, to consider the radial
derivative of the scalar profile and demand its continuity
across the horizons,
\begin{align}
\partial_r \Phi(t, r) ={}&
    \frac{H^2r^2 
    \mathcal{Q} \sqrt{1 \!-\! F(r)}
    +
    4\alpha F'(r) \big[ 1 \!-\! F(r) \big]
	-
	\mathcal{C}}{r^2F(r)}
    \, .
\label{drPhi}
\end{align}
This yields two linear conditions on the parameters,
\begin{align}
    (H r_{\scr \rm BH})^2 \mathcal{Q}
    ={}&
    \mathcal{C}
    -
    4\alpha F'(r_{\scr \rm BH})
    \, ,
\\
    (H r_{\scr \rm C})^2 \mathcal{Q}
    ={}&
	\mathcal{C}
    -
    4\alpha F'(r_{\scr \rm C})
    \, ,
\end{align}
and these conditions uniquely determine~$\mathcal{Q}$ 
and~$\mathcal{C}$,
\begin{align}
\mathcal{Q}
    ={}&
    4\alpha \frac{ F'(r_{\scr \rm BH}) 
        - F'(r_{\scr \rm C}) }
            {H^2 (r_{\scr \rm C}^2 - r_{\scr \rm BH}^2) }
    \, ,
\\
\mathcal{C}
    ={}&
    4\alpha \frac{ r_{\scr \rm C}^2 F'(r_{\scr \rm BH})
        - r_{\scr \rm BH}^2 F'(r_{\scr \rm C}) }
            {r_{\scr \rm C}^2 - r_{\scr \rm BH}^2}
    \, .
\end{align}
This is the value for the constants obtained 
in~\cite{Babichev:2024txe} in the static coordinates, where the 
degeneracy can be broken by appealing to the finiteness of the 
shift-charge current, which comes with its own 
subtleties~\cite{Creminelli:2020lxn}.
In the limit~$H\RS\!\ll\!1$ that we are working in, these
two constants are the same, up to small corrections,
$\mathcal{Q} \!=\! 4\alpha/\RS [1 \!+\! \mathcal{O}(H\RS)] \!=\! \mathcal{C}$.

With this choice, 
the radial derivative of the scalar field is
continuous across both horizons. Although Eq.~(\ref{drPhi}) can
be integrated explicitly, the resulting expression is cumbersome
and will not be needed here. On superhorizon scales, the
derivative simplifies to
\begin{equation}
\partial_r \Phi(t,r) \xrightarrow{r \gg r_{\scr \rm C}}
    \frac{H}{r} \big( 8\alpha H - \mathcal{Q} \big)
    \, ,
\end{equation}
which integrates straightforwardly to
\begin{equation}
\Phi(t,r) \xrightarrow{r \gg r_{\scr \rm C}}
    \mathcal{Q} H^2 t
    +
    \big( 8\alpha H - \mathcal{Q} \big) H \ln(Hr)
    \, .
\end{equation}
This has precisely the same structure as the superhorizon
behavior of the de Sitter point-source
profile~(\ref{dSasymptotic}), upon identifying the effective
source strength as~$\beta\!=\! \mathcal{Q} \!-\! 8\alpha H \!\approx\! \mathcal{C}$. This agreement shows that 
the black hole sources the same superhorizon scalar profile 
as a point source in de Sitter, with an amplitude fixed by 
the local black-hole physics.

This result supports the interpretation of the scalar-hair tail 
in terms of the physical mechanism outlined in 
Sec.~\ref{sec: Scalar point source in de Sitter spacetime}. 
The black hole perturbs the scalar field locally, after which 
its superhorizon evolution is governed by the cosmological 
dynamics of a massless, minimally coupled scalar field.

\subsection{Backreaction}
\begin{figure}[h!]
\center
\includegraphics[width=8.5cm]{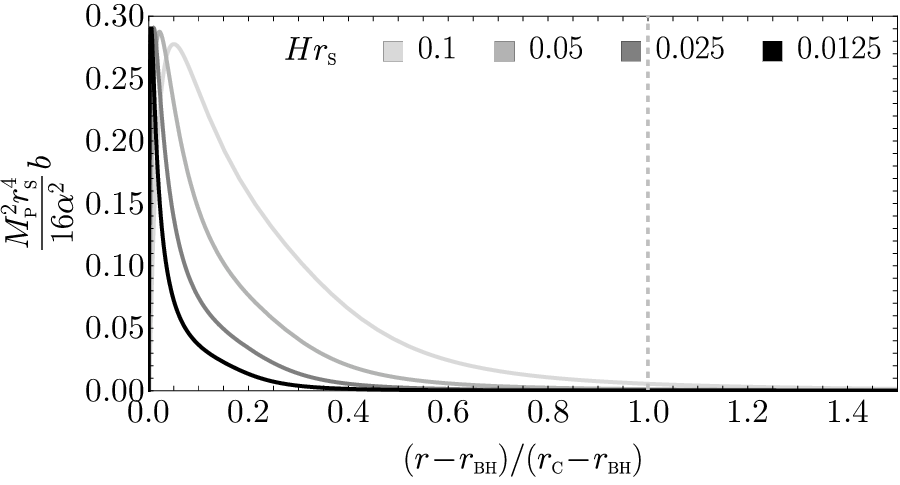}
\vskip-3mm
\caption{Strength of the backreaction of the scalar field on
the Schwarzschild-de Sitter background. Backreaction first
becomes important on subhorizon scales, close to the black
hole, and decreases substantially before the cosmological
horizon.}
\label{SdSBackreactionPlot}
\end{figure}

We evaluate the invariant backreaction diagnostic by comparing
the Kretschmann invariant,
\begin{equation}
\mathcal{K} = \frac{12 r_{\scr \rm S}^2}{r^6} + 24 H^4 
    = \mathcal{G} \, ,
\end{equation}
to the trace of the energy-momentum tensor,
\begin{equation}
\mathcal{T} = 2X + 24\alpha^2 H^2 \mathcal{G} \, .
\end{equation}
Its radial dependence is shown in
Fig.~\ref{SdSBackreactionPlot}.

In the regime where
\begin{equation}
\Big( \frac{4\alpha}{r_{\scr \rm S}} \Big)^{\!2} \ll
    (\MP r_{\scr \rm S})^2
    \, ,
\label{SdShierarchy}
\end{equation}
the backreaction can be neglected everywhere, both on
subhorizon and superhorizon scales, and the test-field
approximation provides a good description of the scalar hair.
When the hierarchy~(\ref{SdShierarchy}) is not satisfied,
backreaction must instead be taken into account. As is clear
from Fig.~\ref{SdSBackreactionPlot}, it becomes important first
on subhorizon scales. Thus, the first breakdown of the
test-field approximation occurs near the black hole, while the
superhorizon regime remains under control. This indicates that
the relevant nonlinear problem is localized and should be
resolved there before drawing conclusions about the
cosmological-scale profile.

\subsection{Scalar monopole radiation}
The parallel between the Schwarzschild-de Sitter spacetime and
the point source in de Sitter extends also to scalar monopole
radiation. In the region between the two horizons, Kottler
coordinates provide a natural framework for describing the
associated stationary energy flux.

The total energy inflow through a sphere of radius~$r$ is again given by
$\dot{M} \!=\! 4\pi r^2 T^r{}_\tau$.
Computing this for our
system gives the constant result,\footnote{This computation
is facilitated by {\it Cadabra}~\cite{Peeters:2007wn,Peeters:2006kp,Peeters:2018dyg}.}
\begin{equation}
\dot{M} = - 4 \pi \mathcal{C} \mathcal{Q} H^2 \, ,
\end{equation}
where again the negative sign indicates an outward energy flux. 
Up to small corrections of
order~$\mathcal{O}(H\RS)$, this result matches the one
obtained for the point scalar source in
de Sitter~(\ref{ACCdS}), upon the same identification of
parameters,~$\beta\!=\!\mathcal{Q} \!-\! 8\alpha H \!\approx \! \mathcal{C}$,
used above to match the asymptotic superhorizon scalar
profiles.

This result is consistent with the behavior of a scalar point
charge in de Sitter spacetime, where the outward flux is
interpreted as monopole scalar radiation. The
Schwarzschild-de Sitter solution therefore admits the same
interpretation; the time-dependent scalar configuration carries
a steady outward energy flux. In this sense, the superhorizon
scalar hair cannot be sustained indefinitely with an arbitrarily
large energy content, since the associated flux is constrained
by the finite mass of the black hole.

For the test-field approximation to remain applicable, the
black hole should not evolve appreciably during a Hubble time,
namely~$|\dot{M}|/H \!\ll\! M$. This condition is equivalent to
\begin{equation}
\Big( \frac{4\alpha}{r_{\scr \rm S}} \Big)^{\!2}
    \ll
    \frac{(\MP r_{\scr \rm S})^2}{H r_{\scr \rm S} }
    \, .
\label{AccretionHierarchy}
\end{equation}
Because we assume~$H r_{\scr \rm S} \!\ll\! 1$, this condition
is weaker than the one obtained for the absence of backreaction
in~(\ref{SdShierarchy}). The energy flux therefore does not set
the leading limitation on the test-field approximation; by the
time it would become relevant, scalar backreaction would
already have become important. This mirrors the asymptotically
flat case, where the energy flux is absent but the backreaction due
to scalar hair can nevertheless become appreciable.

This does not alter our main conclusion about the regime of
validity of the test-field approximation. As the black hole
presumably slowly loses mass, the system is driven toward the 
regime in which local scalar backreaction becomes important. 
Thus, long before the ultimate late-time evolution could be inferred 
from the test-field solution alone, the perturbative treatment would
already cease to be trustworthy near the black hole.
Whether this would terminate the scalar monopole radiation or not
remains an open question.

\section{Discussion}
\label{sec: Discussion}

The results presented here show that the logarithmically growing
scalar-hair profile on cosmological scales in the linearly
coupled shift-symmetric scalar-Gauss-Bonnet theory should not
be interpreted as an inconsistency of the solution or as an
obstruction associated with the cosmological horizon. Rather,
it is the expected large-distance behavior of a massless,
minimally coupled scalar field sourced by a localized object in
an expanding de Sitter background. In this respect, the
cosmological profile is not a special feature of black holes,
but a manifestation of the underlying scalar-field dynamics.

This conclusion is clarified by comparing the three systems
studied in this work. The Schwarzschild analysis isolates the
local scalarization mechanism that fixes the effective scalar
charge of the black hole on subhorizon scales. The point-source
solution in de Sitter then isolates the cosmological evolution
of a localized scalar charge and makes the origin of the
logarithmic superhorizon profile transparent.

Finally, the Schwarzschild-de Sitter solution shows that these
two elements coexist in a single geometry; once the scalar field
is described in coordinates that remain regular across both the
black-hole and cosmological horizons, the black hole reproduces
the same asymptotic structure as the de Sitter point source. In
this sense, the asymptotic scalar profile of the black hole is
the cosmological continuation of locally sourced scalar hair.

The parallel between the Schwarzschild-de Sitter case and
the point source in de Sitter can be understood directly from
the scalar equation of motion. In both cases, the scalar field
behaves as a massless, minimally coupled field sourced by an
effective localized charge: in one case this charge is provided
explicitly by a point source, while in the other it is induced
by the geometry through the Gauss-Bonnet invariant. The latter
may be viewed heuristically as a superposition of point sources
spread through the localized subhorizon region. From the
point of view of the superhorizon scalar profile, this
distinction is largely irrelevant; once the local source fixes
the total scalar charge, the subsequent large-distance
evolution is governed by the same cosmological dynamics.

A related lesson is methodological. The Schwarzschild-de Sitter
problem is most naturally analyzed in Painlev\'{e}-Gullstrand
coordinates, which remain regular at both horizons. In these
coordinates the physically relevant branch is selected
unambiguously by demanding continuity of the radial derivative
of the scalar profile across the two horizons, and the resulting
superhorizon behavior can be read off directly. This is
important because Kottler coordinates, used in earlier
analyses, cover only the static patch and it is not immediately
clear how the scalar hair profile extends beyond it. The coordinates
that we employ, on the other hand, make clear what the long-distance
behavior of the scalar profile is and how it connects to the analogous
scalar profile sourced by point scalar charges.

The point-source analysis also provides a simple physical
picture for the origin of the cosmological profile. A localized
source first excites scalar perturbations on subhorizon scales.
Cosmological expansion then stretches those perturbations to
superhorizon scales and amplifies them in the familiar way for a
massless, minimally coupled scalar field. The momentum-space
spectrum in Fig.~\ref{SpectrumFigure} makes this mechanism
particularly clear: only modes that were initially subhorizon
become populated, and their subsequent evolution gives rise to a
scale-invariant superhorizon spectrum. The same mechanism 
controls the superhorizon profile of the black-hole solution,
while the local black-hole physics enters only through the
effective scalar charge that fixes its overall amplitude.

Our backreaction analysis supports the same interpretation. In
all examples studied here, the first regime in which the
test-field approximation is expected to fail occurs on
subhorizon scales, near the localized source, rather than on
cosmological scales. The dominant nonlinear problem is
therefore not the existence of the logarithmic profile itself,
but the behavior of the scalar configuration in the vicinity of
the black hole. If backreaction becomes important, its primary
effect should be to modify the effective local source that feeds
the cosmological profile, rather than to eliminate the
large-distance behavior generated by cosmological evolution.

Perhaps the most striking ingredient emerging from the analysis is 
the presence of a steady outward energy flux associated with the
time-dependent scalar configuration. This occurs already for a
point scalar charge in de Sitter spacetime, where it is
naturally interpreted as monopole scalar radiation. The
Schwarzschild-de Sitter solution exhibits the same effect up to
small corrections in~$H r_{\scr \rm S}$.

Even though this flux does not set the leading limitation
on the test-field approximation in the regime
$H r_{\scr \rm S} \!\ll\! 1$, it does provide a new physical 
picture of the system where the scalarized black hole is embedded
in de Sitter space asymptotics. Even though the energy-momentum tensor
of such a configuration is static, it encodes an outward energy flux 
that prevents the system from being considered stationary. This
flux signals that the scalarized black hole embedded in cosmology does
not necessarily have to resemble the scalarized black hole with flat
asymptotics. The outward flux seems to push the black hole towards 
smaller Schwarzschild radii, bringing it steadily closer to the 
regime where backreaction becomes important.

Although self-consistent cosmological solutions have not been
constructed here, the test-field results strongly suggest that
the cosmological-scale profile should persist even when the
near-source region becomes nonlinear. A fully nonlinear
black-hole configuration may differ substantially from the
perturbative solution near the horizon, but as long as it
remains a localized source of scalar charge, its effect at
large distances should still be governed by the same
cosmological mechanism identified in this work. This also
clarifies why earlier attempts to construct self-consistent
black-hole solutions with de Sitter asymptotics under the
assumption of a static scalar profile encounter difficulty; the
cosmological dynamics of a massless, minimally coupled scalar
field generically introduces time dependence into the scalar
hair, together with an associated energy flux. Currently it is
not clear whether the black hole keeps radiating scalar 
monopole radiation indefinitely, or whether there is an
equilibrium end point to this process where the scalar charged
is somehow screened. It is also interesting to note that modifications 
of the scalar kinetic term can turn the cosmological hair from secondary to 
primary~\cite{Lara:2025hqh}, though it is not completely clear to
which degree this distinction applies to situations where hair is not stationary.

\vspace{-2em}
\section*{Acknowledgments}
\noindent
We are indebted to Ignacy Sawicki for discussions on 
the subject, and for the critical reading of the manuscript.
We also thank José Luis Blázquez-Salcedo and Alexander Vikman 
for useful comments. D.G. was supported by the Czech Science 
Foundation (GA\v{C}R) Grant No. 24-13079S. D.J.G. acknowledges support 
from the Comunidad de Madrid under predoctoral Contract No. PIPF-2023/TEC-29931, and thanks FZU's Department of Cosmology 
and Gravitational Physics for their kind hospitality.



\end{document}